\documentclass[cits]{PoS}
\usepackage{clshan-math,clshan-dm-dd}
\newcommand{\imageswitch} [2] {#2}
\def \lsim {\:\raisebox{-0.7ex}{$\stackrel{\textstyle<}{\sim}$}\:}
\def \gsim {\:\raisebox{-0.7ex}{$\stackrel{\textstyle>}{\sim}$}\:}
\def \ignore#1 {}
\title{Effects of Residue Background Events \\
       in Direct Detection Experiments on   \\
       Determining Properties of Halo Dark Matter}
\ShortTitle{Background effects
            in direct detection experiments on
            determining halo DM properties}
\author{\speaker{Chung-Lin Shan} \\ 
        Department of Physics, National Cheng Kung University,
        Tainan City 70101, Taiwan, R.O.C.                     \\
        Physics Division,
        National Center for Theoretical Sciences,
        Hsinchu City 30013, Taiwan, R.O.C.                    \\
        E-mail: \email{clshan@mail.ncku.edu.tw}}
\abstract{
 We reexamine the model--independent data analysis methods
 for extracting properties of
 Weakly Interacting Massive Particles (WIMPs)
 by using data (measured recoil energies) from
 direct Dark Matter detection experiments directly and,
 as a more realistic study,
 consider a small fraction of residue background events,
 which pass all discrimination criteria and
 then mix with other real WIMP--induced signals
 in the analyzed data sets.
 In this talk,
 the effects of residue backgrounds
 on the determination of
 the mass of halo Dark Matter particle
 as well as
 on the reconstruction of its one--dimensional
 velocity distribution function
 will be discussed.
}

\FullConference{Identification of dark matter 2010 \\
                July 26-30, 2010 \\
                Montpellier, France}
\begin{document}
\twocolumn
\section{Introduction}
 In our earlier work on the development of
 model--independent data analysis methods
 for extracting properties of
 Weakly Interacting Massive Particles (WIMPs)
 by using measured recoil energies
 from direct Dark Matter detection experiments directly
 \cite{DMDDf1v, DMDDmchi, DMDD-sigma},
 it was assumed that
 the analyzed data sets are background--free,
 i.e., all events are WIMP signals.
 Active background discrimination techniques
 should make this condition possible.
 For example,
 by using the ratio of the ionization to recoil energy,
 the so--called ``ionization yield'',
 combined with the ``phonon pulse timing parameter'',
 the CDMS-II collaboration claimed that
 electron recoil events can be rejected event--by--event
 with a misidentification fraction of $< 10^{-6}$.
 \cite{Ahmed09b}
%
 The CRESST collaboration demonstrated also that
 the pulse shape discrimination (PSD) technique
 can distinguish WIMP--induced nuclear recoils
 from those induced by backgrounds
 by means of inserting a scintillating foil,
 which causes some additional scintillation light
 for events induced by $\alpha$-decay of $\rmXA{Po}{210}$
 and thus shifts the pulse shapes of these events
 faster than pulses induced by WIMP interactions in the crystal
 \cite{CRESST-bg}.
\footnote{
 More details
 about background discrimination techniques and status
 see also e.g.,
 Refs.~\cite{bg-papers}.
}

 However,
 as the most important issue in all
 underground experiments,
 possible residue background events
 which pass all discrimination criteria and
 then mix with other real WIMP--induced events in our data sets
 should also be considered.
 Therefore,
 as a more realistic study,
 we take into account
 small fractions of residue background events
 mixed in experimental data sets
 and want to study
 how well the model--independent methods
 could extract the {\em input} \\ WIMP properties
 by using these ``impure'' data sets
 and how ``dirty'' these data sets could be
 to be still useful.

 In this article,
 I focus on two properties of halo WIMPs:
 their mass $\mchi$ and
 one--dimensional velocity distribution function $f_1(v)$.
 More detailed discussions
 can be found in Refs.~\cite{DMDDbg-mchi, DMDDbg-f1v}.
\section{Effects of residue background events}
 In our numerical simulations
 based on the Monte Carlo method,
 while the shifted Maxwellian velocity distribution
 \cite{SUSYDM96, DMDDf1v}
 with the standard values of
 the Sun's orbital velocity
 and the Earth's velocity
 in the Galactic frame:
 $v_0 \simeq 220~{\rm km/s}$
 and
 $\ve = 1.05 \~ v_0$,
 and the Woods--Saxon form
 for the elastic nuclear form factor
 for the spin--independent (SI) WIMP--nucleus interaction
 \cite{Engel91, SUSYDM96}
 have been use for generating WIMP--induced signals,
 a {\em target--dependent exponential} form
 for residue background events has been introduced
 \cite{DMDDbg-mchi}:
\beq
   \aDd{R}{Q}_{\rm bg, ex}
 = \exp\abrac{-\frac{Q /{\rm keV}}{A^{0.6}}}
\~.
\label{eqn:dRdQ_bg_ex}
\eeq
 Here $Q$ is the recoil energy,
 $A$ is the atomic mass number of the target nucleus.
 The power index of $A$, 0.6, is an empirical constant,
 which has been chosen so that
 the exponential background spectrum is
 somehow {\em similar to},
 but still {\em different from}
 the expected recoil spectrum of the target nucleus
 (see Figs.~\ref{fig:dRdQ-bg-ex-Ge-000-100-20});
 otherwise,
 there is in practice no difference between
 the WIMP scattering and background spectra.
 Note that,
 the atomic mass number $A$
 has been used here
 just as the simplest, unique characteristic parameter
 in the 
 analytic form (\ref{eqn:dRdQ_bg_ex})
 for defining the residue background spectrum
 for {\em different} target nuclei.
 It does {\em not} mean that
 the (superposition of the real) background spectra
 would depend simply/primarily on $A$ or
 on the mass of the target nucleus, $\mN$.

 Note also that,
 firstly,
 the exponential form (\ref{eqn:dRdQ_bg_ex})
 for residue background spectrum
 is rather naive;
 however,
 since we consider here
 {\em only a few (tens) residue} background events
 induced by perhaps {\em two or more} different sources,
 pass all discrimination criteria,
 and then mix with other WIMP--induced events
 in our data sets of ${\cal O}(100)$ {\em total} events,
 exact forms of different background spectra
 are actually not very important and
 this exponential spectrum
 should practically not be unrealistic.
 Secondly,
 our model--independent data analysis procedures
 requires only measured recoil energies
 from one or more experimental data sets
 with different target nuclei
 \cite{DMDDf1v, DMDDmchi, DMDD-sigma}.
 Hence,
 for applying these methods to future real direct detection data,
 the prior knowledge about (different) background source(s)
 is {\em not required at all}.

 Moreover,
 the maximal cut--off of the velocity distribution function
 has been set as \\ \mbox{$\vmax = 700$ km/s}.
 The experimental threshold energy
 has been assumed to be negligible
 and the maximal cut--off energy
 is set as 100 keV.
 The background window
 (the possible energy range
  in which residue background events {\em cannot be ignored},
  compared to some other ranges)
 has been assumed to be the same as
 the experimental possible energy range.
 Note here that
 the actual numbers of generated signal and background events
 in each simulated experiment
 are Poisson--distributed around their expectation values
 {\em independently},
 and the total event number in one experiment
 is then the sum of these two numbers;
 both generated signal and background events
 are treated as WIMP signals in our analyses.
 Additionally,
 we assumed that
 all experimental systematic uncertainties
 as well as the uncertainty on
 the measurement of the recoil energy
 could be ignored.
\subsection{On the measured recoil spectrum}
\begin{figure}[p!]
\begin{center}
\imageswitch{}
{\hspace*{-0.75cm}
 \includegraphics[width=8.8cm]{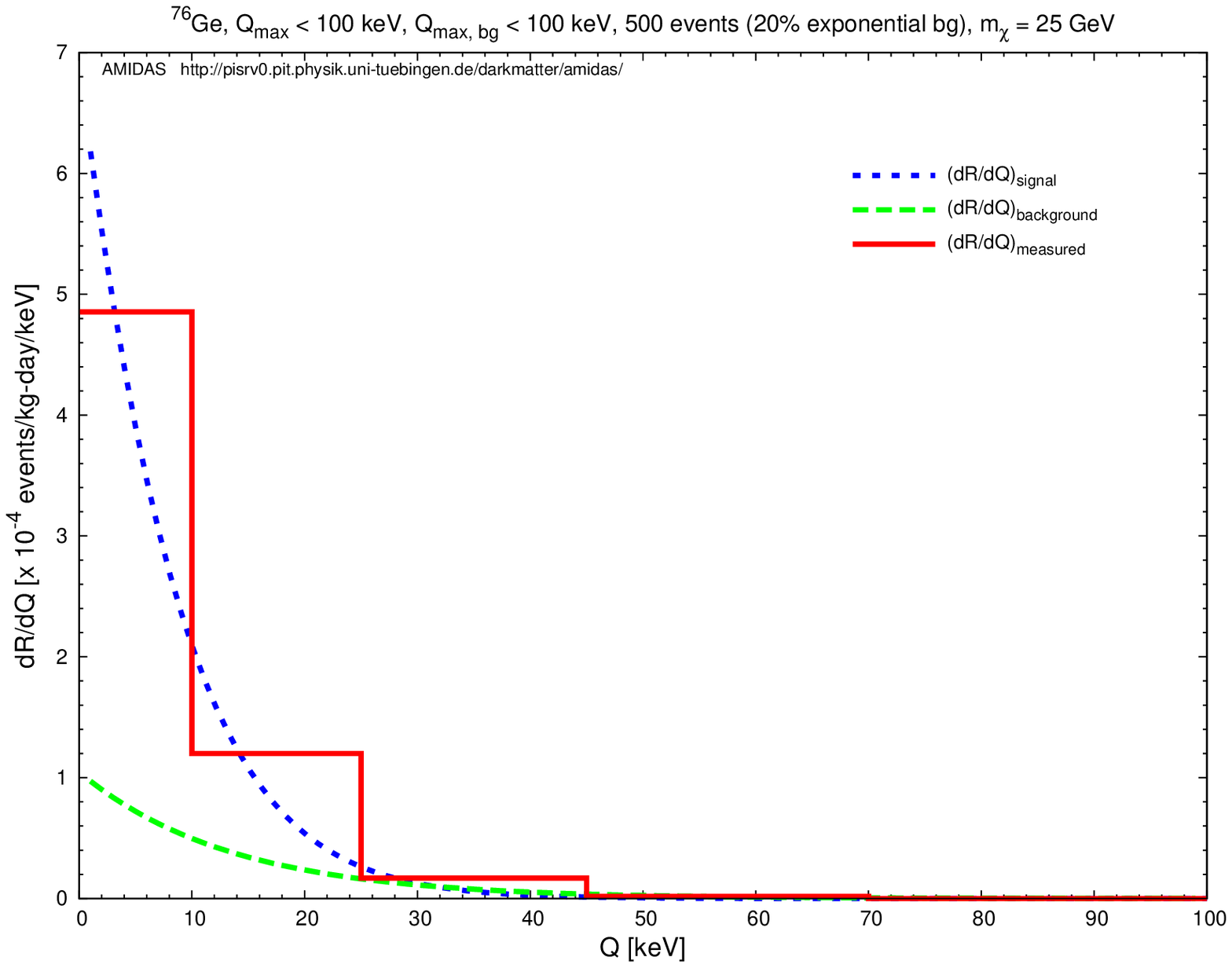} \\ 
 \hspace*{-0.75cm}
 \includegraphics[width=8.8cm]{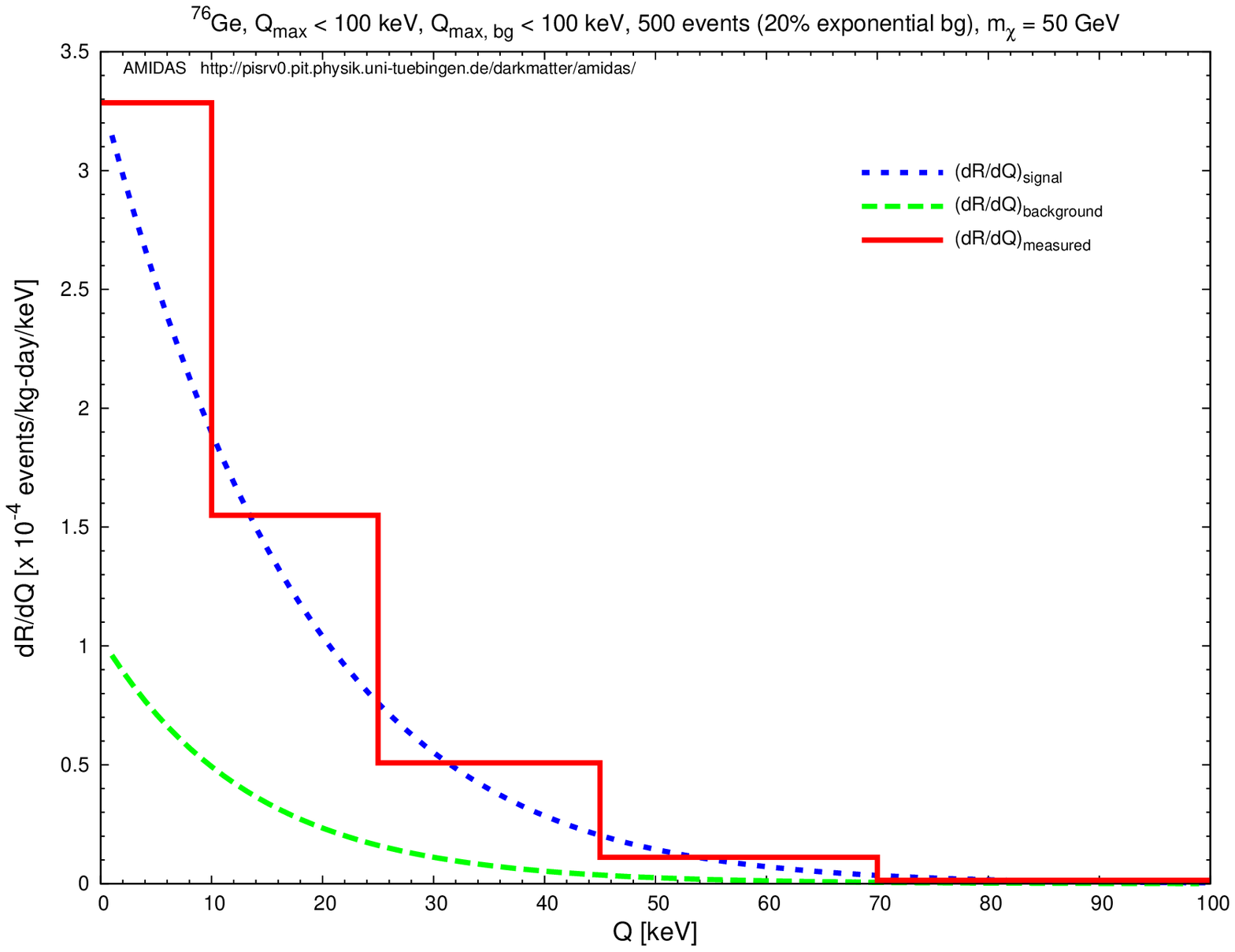} \\ 
 \hspace*{-0.75cm}
 \includegraphics[width=8.8cm]{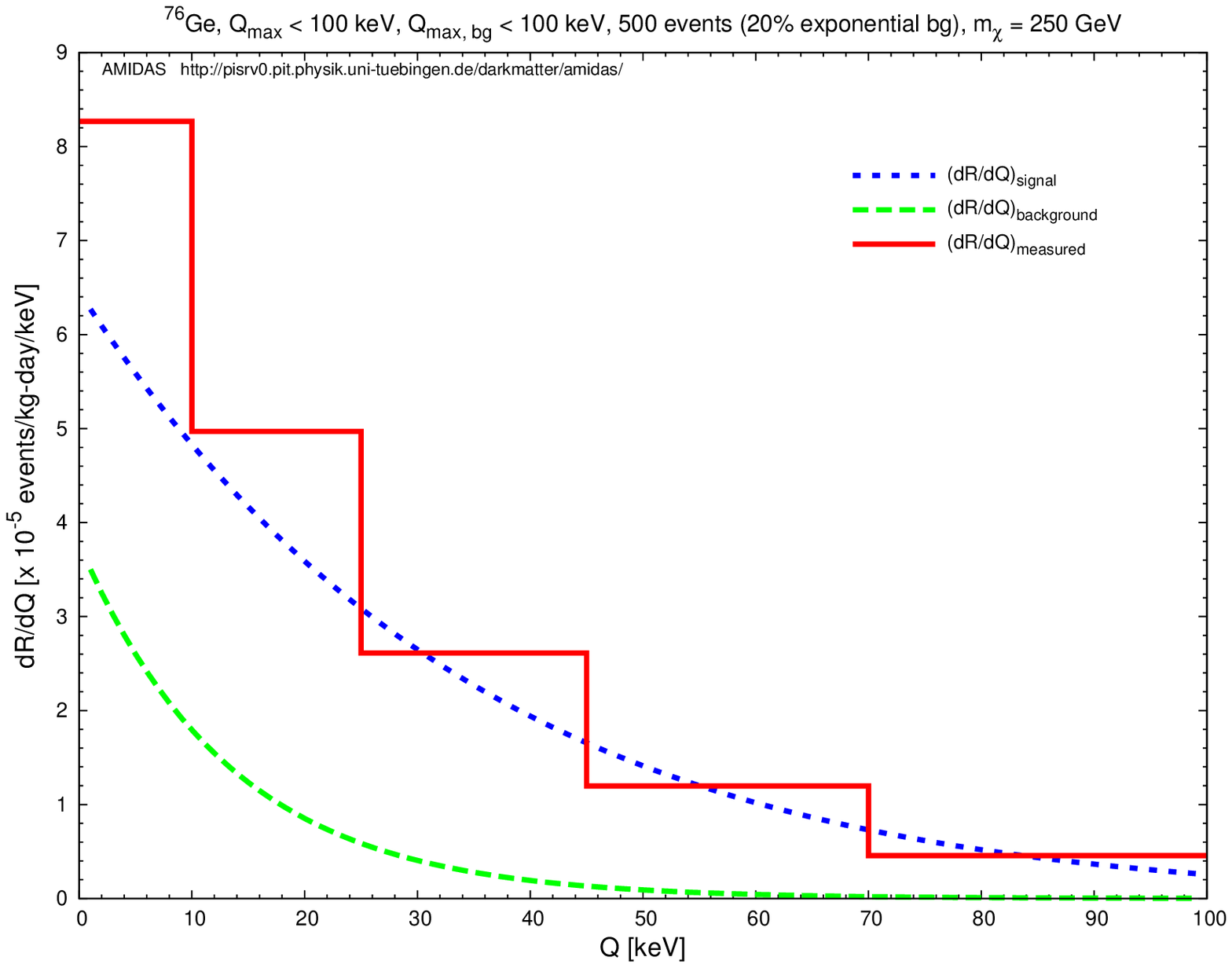} \\ 
}
\vspace{-0.75cm}
\end{center}
\caption{
 Measured energy spectra (solid red histograms)
 for a $\rmXA{Ge}{76}$ target
 with three different WIMP masses:
 25 (top), 50 (middle), and 250 (bottom) GeV.
 The dotted blue curves are
 the elastic WIMP--nucleus scattering spectra,
 whereas
 the dashed green curves are
 the exponential background spectra
 normalized to fit to the chosen background ratio,
 which has been set as 20\% here
 (plots from Ref.~\cite{DMDDbg-mchi}).
}
\label{fig:dRdQ-bg-ex-Ge-000-100-20}
\end{figure}

 In Figs.~\ref{fig:dRdQ-bg-ex-Ge-000-100-20}
 I show measured energy spectra (solid red histograms)
 for a $\rmXA{Ge}{76}$ target
 with three different WIMP masses:
 25 (top), 50 (middle), and 250 (bottom) GeV.
 While the dotted blue curves show
 the elastic WIMP--nucleus scattering spectra,
 the dashed green curves indicate
 the exponential background spectrum
 given in Eq.~(\ref{eqn:dRdQ_bg_ex}),
 which have been normalized so that
 the ratios of the areas under these background spectra
 to those under the (dotted blue) WIMP scattering spectra
 are equal to the background--signal ratio
 in the whole data sets.
 5,000 experiments with 500 total events on average
 in each experiment have been simulated.

 It can be found here that,
 the shape of the WIMP scattering spectrum
 depends highly on the WIMP mass:
 for light WIMPs ($\mchi~\lsim~50$ GeV),
 the recoil spectra drop sharply with increasing recoil energies,
 while for heavy WIMPs ($\mchi~\gsim~100$ GeV),
 the spectra become flatter.
 In contrast,
 the exponential background spectra shown here
 depend only on the target mass
 and are rather {\em flatter}/{\em sharper}
 for {\em light}/{\em heavy} WIMP masses
 compared to the WIMP scattering spectra.
 This means that,
 once input WIMPs are {\em light}/{\em heavy},
 background events would contribute relatively more to
 {\em high}/{\em low} energy ranges,
 and, consequently,
 the measured energy spectra
 would mimic scattering spectra
 induced by {\em heavier}/ {\em lighter} WIMPs.
\subsection{On determining the WIMP mass}
\begin{figure}[t!]
\begin{center}
\imageswitch{}
{\hspace*{-0.75cm}
 \includegraphics[width=8.8cm]{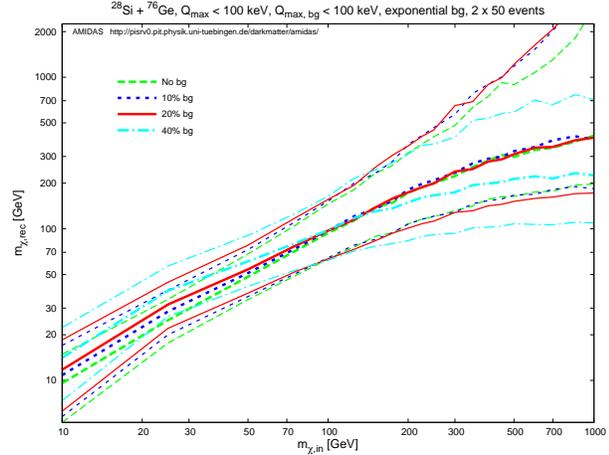}  \\
}
\vspace{-0.75cm}
\end{center}
\caption{
 The reconstructed WIMP masses
 as functions of the input WIMP mass.
 $\rmXA{Si}{28}$ and $\rmXA{Ge}{76}$
 have been chosen as two target nuclei.
 The background ratios shown here
 are no background (dashed green),
 10\% (long--dotted blue),
 20\% (solid red),
 and 40\% (dash--dotted cyan)
 background events in the analyzed data sets.
 Each experiment contains 50 total events
 on average.
 Other parameters are as
 in Figs.~1 
 (plot from Ref.~\cite{DMDDbg-mchi}).
\vspace{-0.25cm}
}
\label{fig:mchi-SiGe-ex-000-100-050}
\end{figure}

 Fig.~\ref{fig:mchi-SiGe-ex-000-100-050}
 show the {\em median} values of
 the reconstructed WIMP mass
 and the lower and upper bounds of
 the 1$\sigma$ statistical uncertainty
 by means of the model--independent procedure introduced
 in Refs.~\cite{DMDDmchi}
 with mixed data sets
 from WIMP--induced and background events
 as functions of the input WIMP mass.
 As in Refs.~\cite{DMDDmchi},
 $\rmXA{Si}{28}$ and $\rmXA{Ge}{76}$
 have been chosen as two target nuclei.
 The background ratios shown here
 are no background (dashed green),
 10\% (long--dotted blue),
 20\% (solid red),
 and 40\% (dash--dotted cyan)
 background events in the analyzed data sets.
 2 $\times$ 5,000 experiments
 with 50 total events on average in each experiment
 have been simulated.

 It can be seen here clearly that,
 since
 for {\em light} WIMP masses ($\mchi~\lsim~100$ GeV),
 due to the relatively flatter background spectrum
 (compared to the scattering spectrum induced by WIMPs)
 or, in practice,
 some background sources in high energy ranges,
 the energy spectrum of all recorded events
 would mimic a scattering spectrum induced
 by WIMPs with a relatively {\em heavier} mass,
 the reconstructed WIMP masses
 as well as the statistical uncertainty intervals
 could be {\em overestimated}.
 In contrast,
 for {\em heavy} WIMP masses ($\mchi~\gsim~100$ GeV),
 due to the relatively sharper background spectrum
 or e.g., some electronic noise,
 relatively more background events
 contribute to low energy ranges,
 the energy spectrum of all recorded events
 would thus mimic a scattering spectrum induced
 by WIMPs with a relatively {\em lighter} mass.
 Hence,
 the reconstructed WIMP masses
 as well as the statistical uncertainty intervals
 could be {\em underestimated}.
 Nevertheless,
 Fig.~\ref{fig:mchi-SiGe-ex-000-100-050}
 shows that,
 with $\sim$ 20\% residue background events
 in the analyzed data sets
 of $\sim$ 50 total events,
 the 1$\sigma$ statistical uncertainty band
 can cover the true WIMP mass pretty well;
 if WIMPs are light ($\mchi~\lsim~200$ GeV),
 the maximal acceptable fraction of
 residue background events
 could even be as large as $\sim$ 40\%.
\subsection{On reconstructing the one--dimensional
            WIMP velocity distribution function}
 In this section
 I show the {\em median} values of
 the reconstructed one--dimensional
 velocity distribution function of halo WIMPs%
\footnote{
 Note that,
 since the experimental maximal cut--off energy
 is fixed as 100 keV,
 for heavy input WIMP masses ($\mchi~\gsim~250$ GeV),
 one can reconstruct the velocity distribution function
 only in the velocity range $v~\lsim~300$ km/s.
}
 with its 1$\sigma$ statistical uncertainty
 by means of the model--independent method
 introduced in Ref.~\cite{DMDDf1v}
 with mixed data sets.
 As in Ref.~\cite{DMDDf1v},
 a $\rmXA{Ge}{76}$ nucleus has been chosen
 as our detector target for reconstructing $f_1(v)$;
 while a $\rmXA{Si}{28}$ target
 and a {\em second} $\rmXA{Ge}{76}$ target
 have been used for determining $\mchi$.
 The background ratios shown here
 are no background (dashed green),
 10\% (long--dotted blue),
 and 20\% (solid red)
 background events in the analyzed data set(s).
 (3 $\times$) 5,000 experiments
 with 500 total events on average in each experiment
 have been simulated.
\subsubsection{With a precisely known WIMP mass}
\begin{figure}[p!]
\begin{center}
\imageswitch{}
{\hspace*{-0.75cm}
 \includegraphics[width=8.8cm]{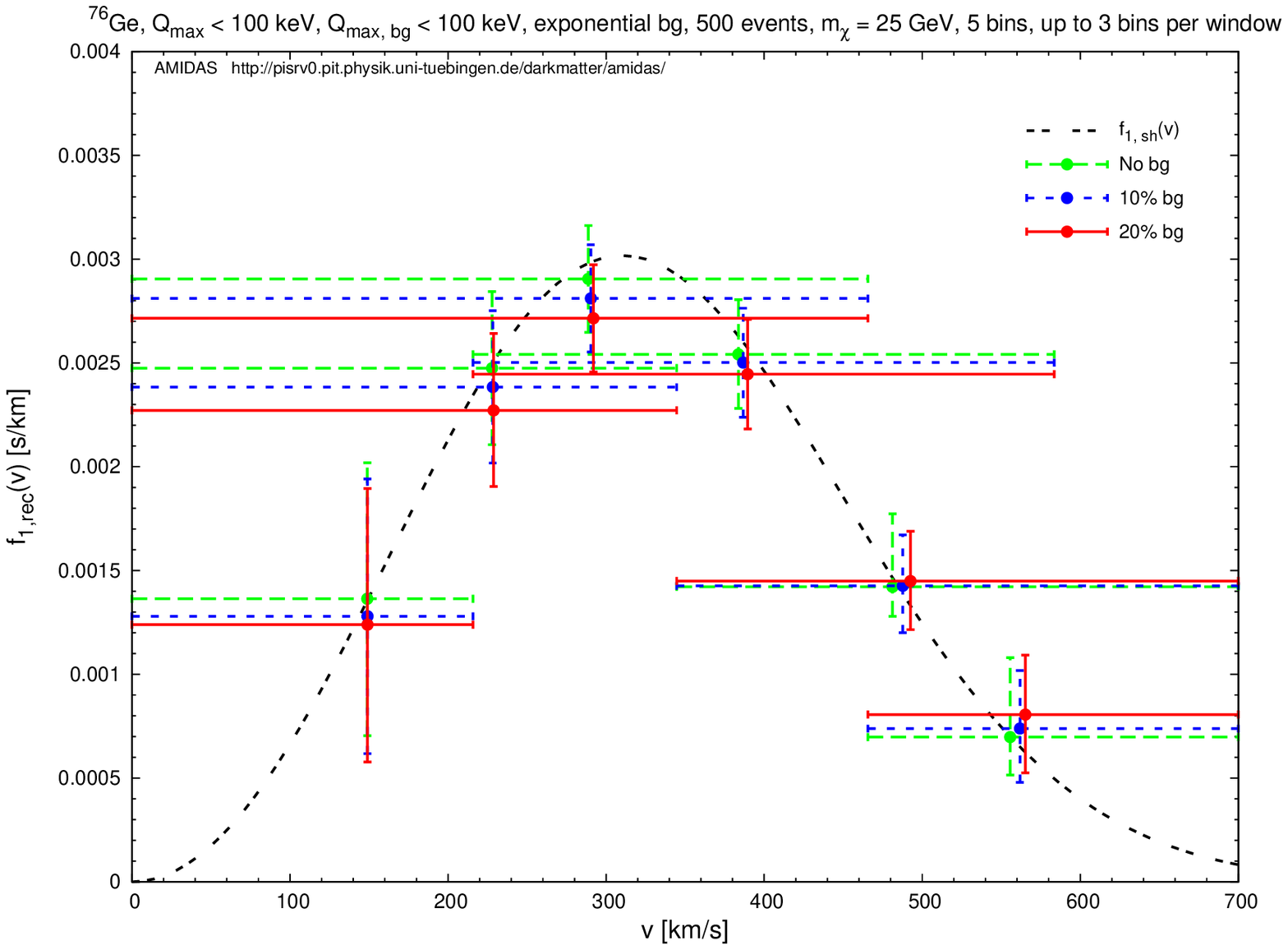} \\ 
 \hspace*{-0.75cm}
 \includegraphics[width=8.8cm]{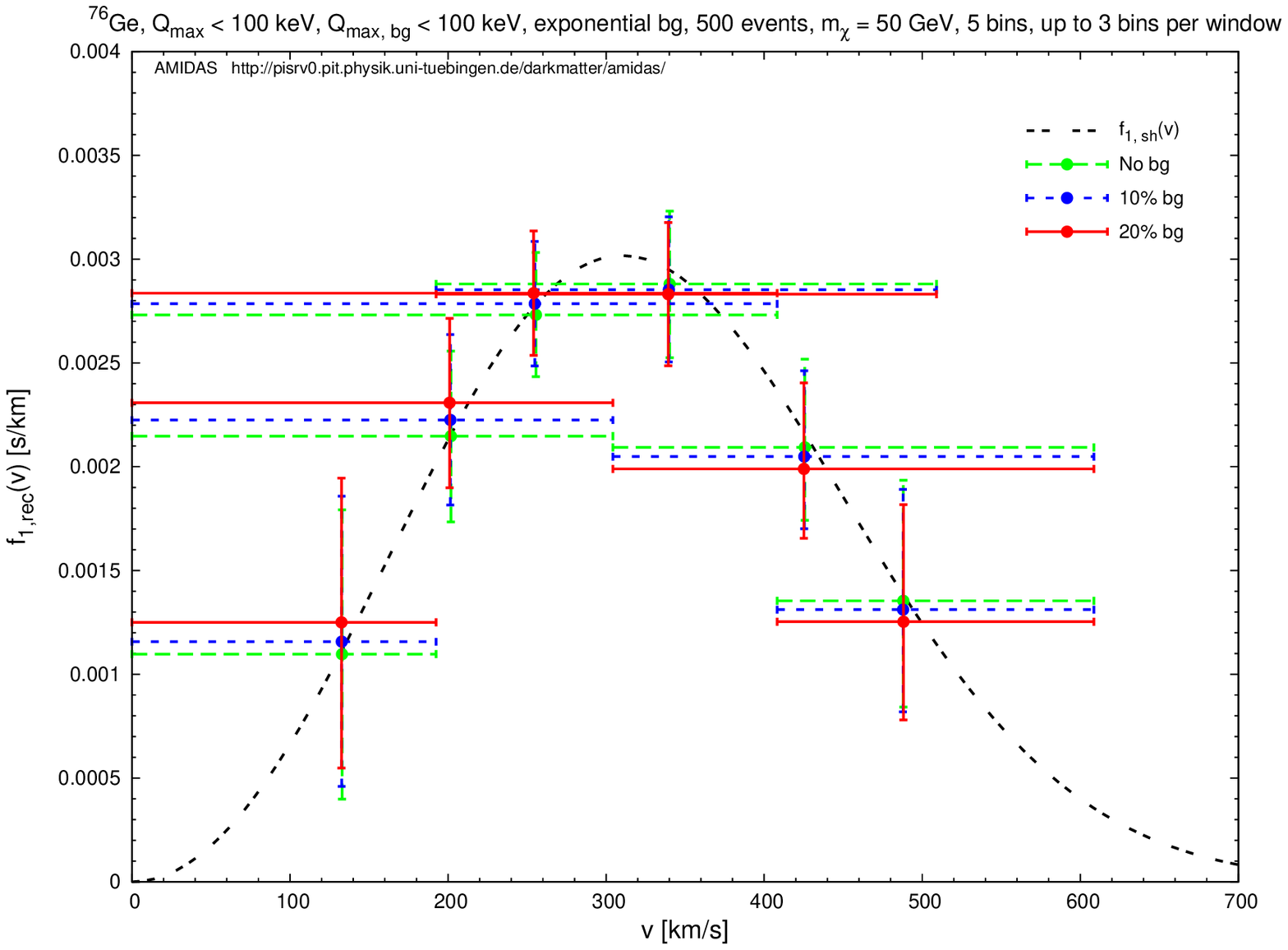} \\ 
 \hspace*{-0.75cm}
 \includegraphics[width=8.8cm]{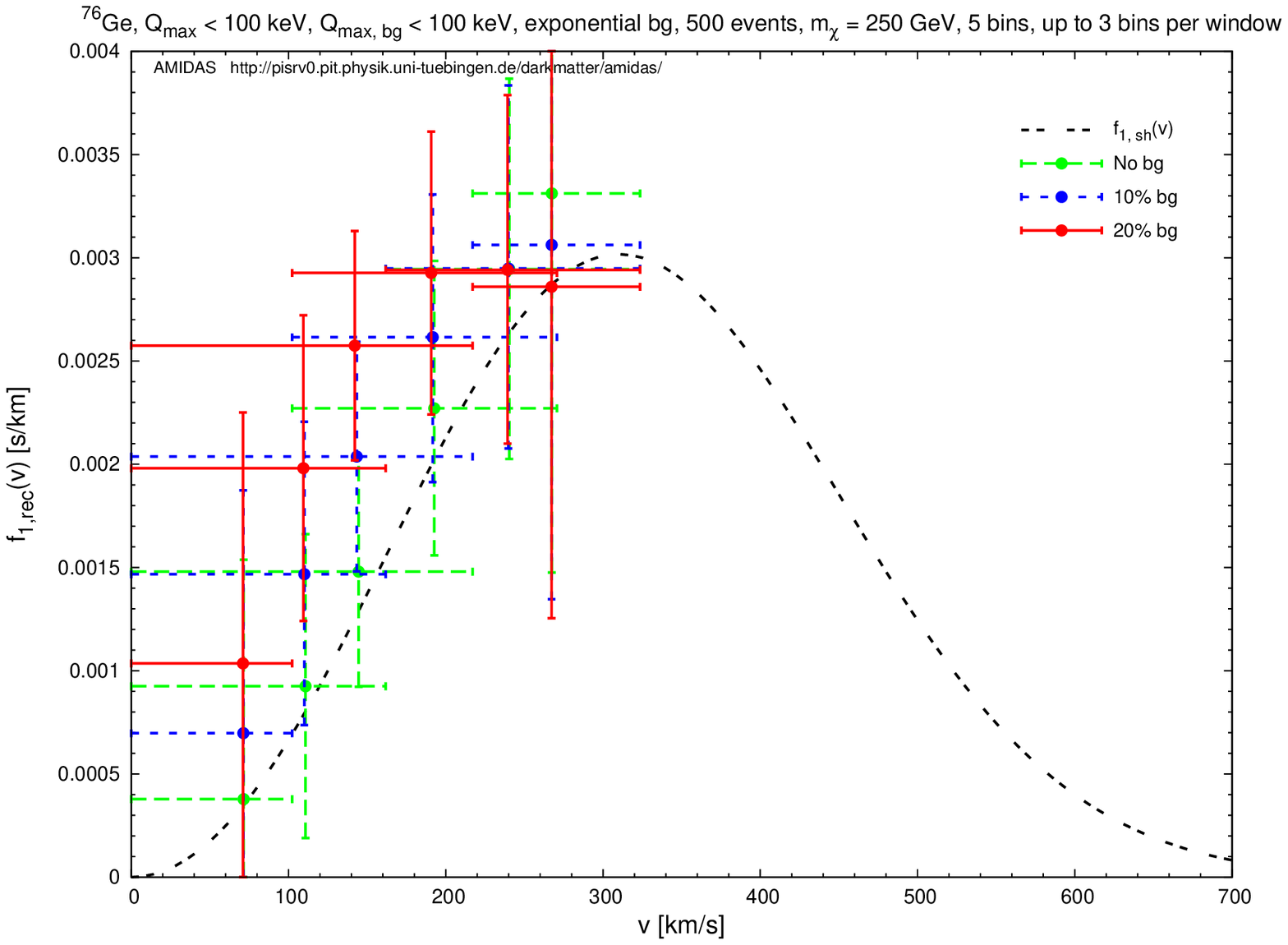} \\ 
}
\vspace{-0.75cm}
\end{center}
\caption{
 The reconstructed one--dimensional WIMP velocity distribution function
 for three different WIMP masses:
 25 (top), 50 (middle), and 250 (bottom) GeV.
 The double--dotted black curves
 are the input shifted Maxwellian velocity distribution.
 The background ratios shown here
 are no background (dashed green),
 10\% (long--dotted blue),
 and 20\% (solid red)
 background events in the analyzed data set.
 Each experiment contains 500 total events
 on average.
 Other parameters are as in Figs.~1 
 (plots from Ref.~\cite{DMDDbg-f1v}).
}
\label{fig:f1v-Ge-ex-000-100-0500}
\end{figure}

 In Figs.~\ref{fig:f1v-Ge-ex-000-100-0500}
 we first assume that
 the required WIMP mass
 for reconstructing $f_1(v)$
 has been known precisely
 with a negligible uncertainty.
 The horizontal bars show
 the sizes of the windows used
 for estimating $f_1(v)$ \cite{DMDDf1v}.
 Five bins have been used and
 up to three bins have been combined to a window,
 in order to collect more events in wider ranges
 and thereby reduce the statistical uncertainty
 on the reconstructed $f_1(v)$.%
\footnote{
 Note that,
 since the neighboring windows overlap,
 the estimates of $f_1(v)$ at adjacent points
 are correlated.
}

 It can be seen that,
 since
 for {\em heavy} WIMP masses ($\mchi~\gsim~100$ GeV),
 the relatively sharper background spectrum
 contributes more events to low energy ranges,
 or, equivalently,
 to low velocity ranges,
 the reconstructed velocity distribution
 would be shifted to {\em lower} velocities.
 For an input WIMP mass of 100 GeV
 and the background ratio of 10\% (20\%),
 the peak of the reconstructed
 velocity distribution function
 could be shifted by $\sim$ 30 (60) km/s.
 In contrast,
 for {\em light} WIMP masses ($\mchi~\lsim~50$ GeV),
 the relatively flatter background spectrum
 contributes more events to high energy/velocity \\ ranges,
 and the reconstructed velocity distribution
 would be shifted to {\em higher} velocities.
 Moreover,
 our simulation results
 indicate that,
 with an $\sim$ 10\% -- 20\% background ratio
 in the analyzed data set of $\sim$ 500 total events,
 one could in principle reconstruct
 the one--dimensional velocity distribution function of halo WIMPs
 with an $\sim -6.5\%$
 (for a 25 GeV WIMP mass, 20\% background events)
 -- $\sim +38\%$
 (for a 250 GeV WIMP mass, 10\% background events)
 deviation.
 If the mass of halo WIMPs is $\cal O$(50 GeV),
 the maximal acceptable background ratio
 could even be as large as $\sim$ 40\%
 with a deviation of only $\sim +14\%$.
\subsubsection{With a reconstructed WIMP mass}
\begin{figure}[t!]
\begin{center}
\imageswitch{}
{\hspace*{-0.75cm}
 \includegraphics[width=8.8cm]{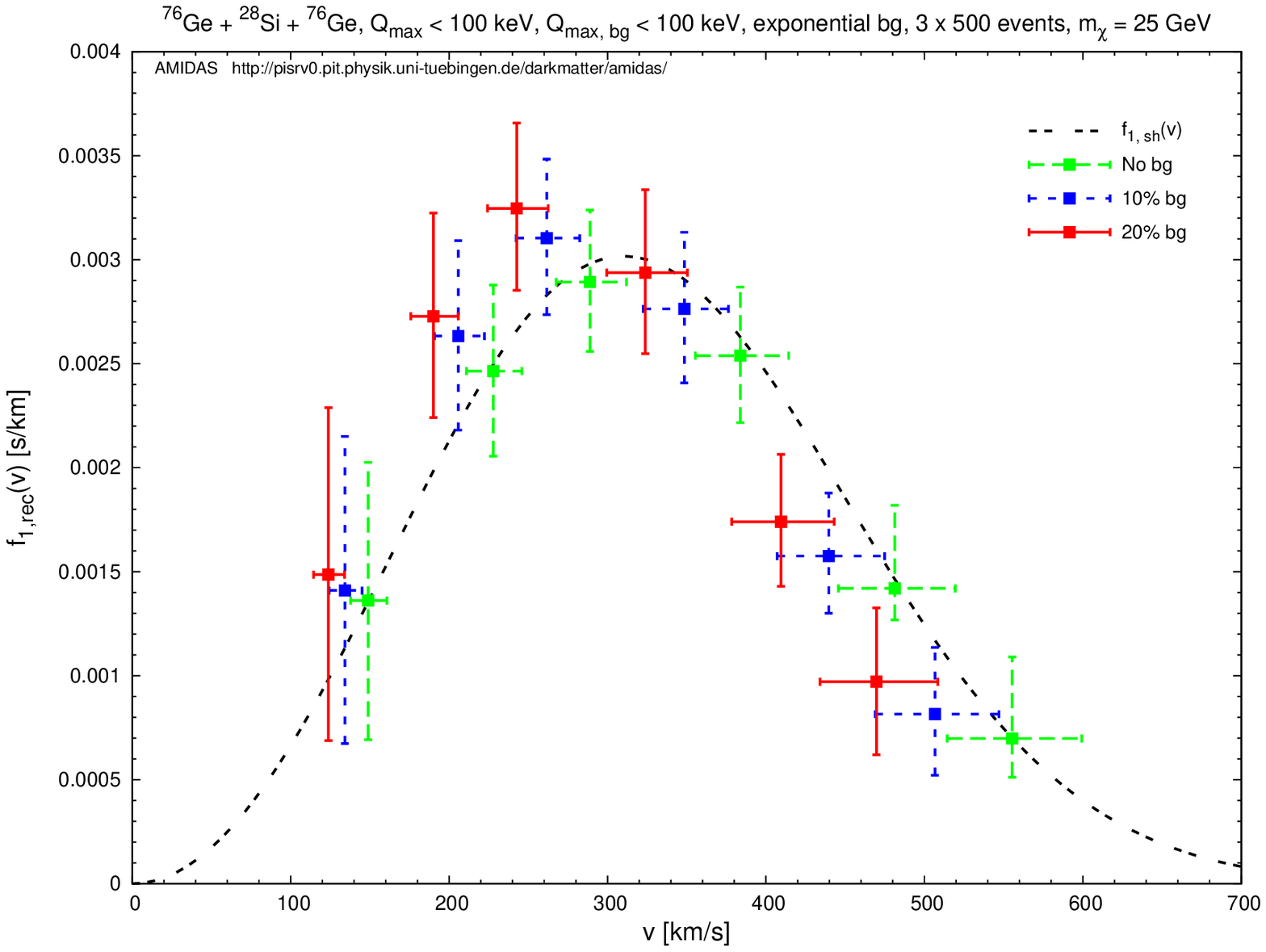} \\ 
 \hspace*{-0.75cm}
 \includegraphics[width=8.8cm]{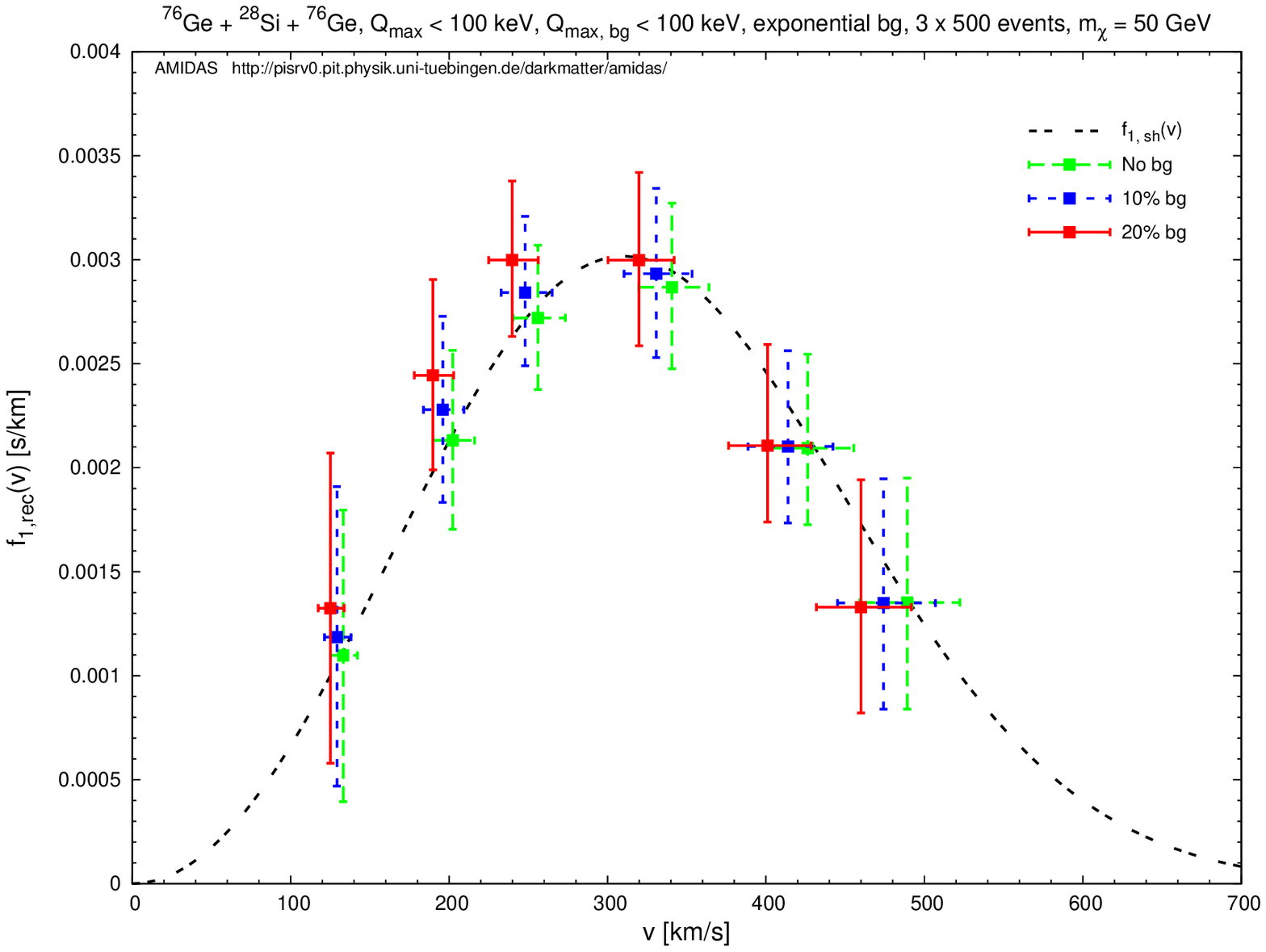} \\ 
 \hspace*{-0.75cm}
 \includegraphics[width=8.8cm]{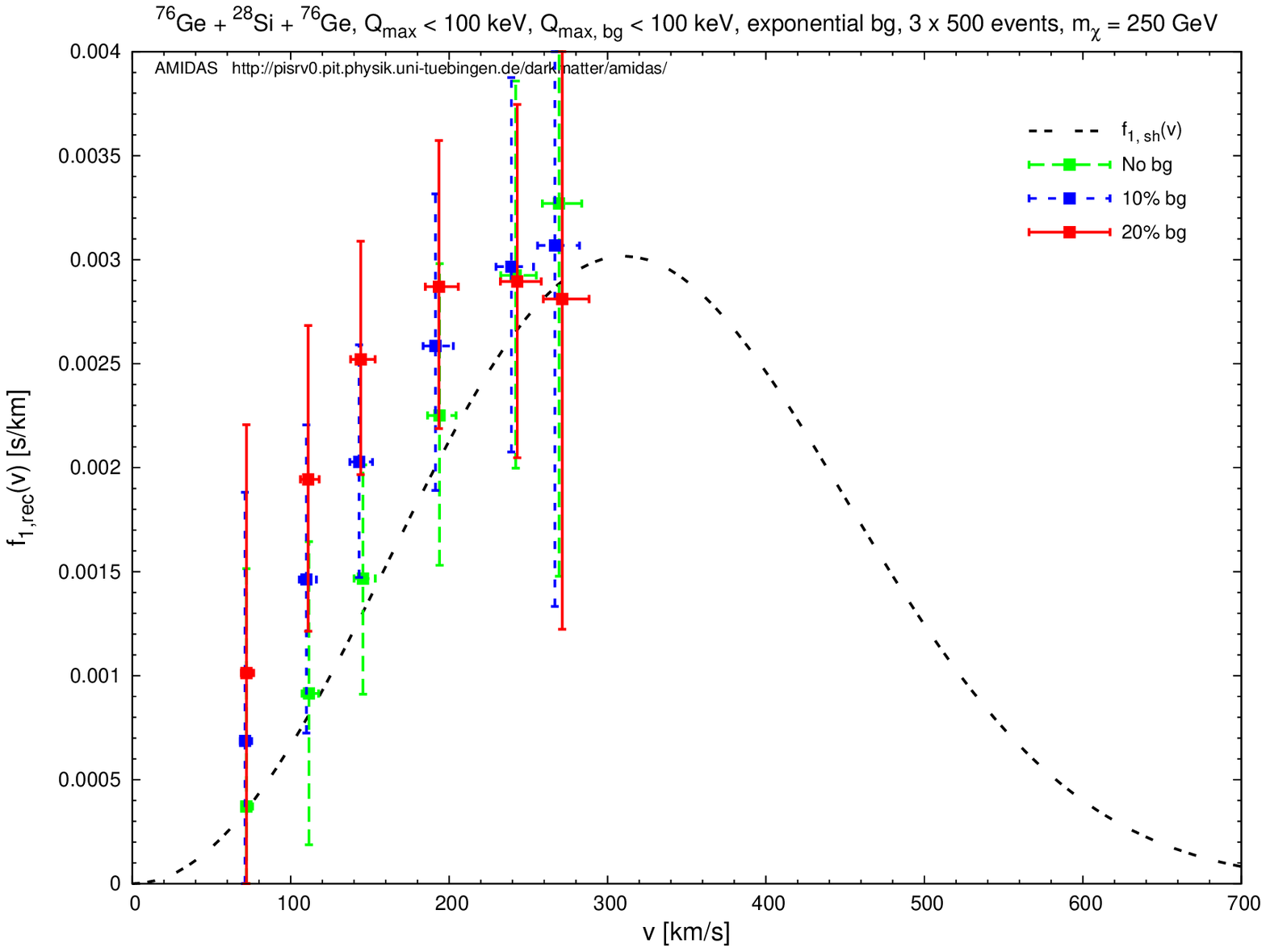} \\ 
}
\vspace{-0.75cm}
\end{center}
\caption{
 As in Figs.~3, 
 except that
 the WIMP masses have been reconstructed
 by means of the procedure introduced
 in Refs.~\cite{DMDDmchi}
 (plots from Ref.~\cite{DMDDbg-f1v}).
\vspace{-0.2cm}
}
\label{fig:f1v-Ge-SiGe-ex-000-100-0500}
\end{figure}

 In Figs.~\ref{fig:f1v-Ge-SiGe-ex-000-100-0500}
 the required WIMP mass
 for reconstructing $f_1(v)$
 has been reconstructed
 with {\em other} direct detection experiments.
 Note that,
 while the vertical bars show
 the 1$\sigma$ statistical uncertainties
 on the reconstructed $f_1(v)$,
 taken into account
 the statistical uncertainty on
 the reconstructed WIMP mass,
 the horizontal bars
 indicate here the 1$\sigma$ statistical uncertainties
 on the estimates of the reconstructed $v$ points,
 due to the uncertainty on the reconstructed $\mchi$.

 It has been found that,
 firstly,
 as shown in Figs.~\ref{fig:mchi-SiGe-ex-000-100-050},
 for an input WIMP mass of 100 GeV,
 the reconstructed mass doesn't differ very much
 from the true value.
 Hence,
 the reconstructed $f_1(v)$
 is approximately the same
 for both cases with the input and reconstructed WIMP masses.
 However,
 for {\em light}/{\em heavy} input masses
 ($\mchi~\lsim/$ $\gsim~100$ GeV),
 the reconstructed $f_1(v)$
 with the reconstructed WIMP mass
 shift to relatively \\ {\em lower}/{\em higher} velocities
 compared to the case
 with the input (true) WIMP mass.
 This effect caused directly by
 the over-/underestimate of the reconstructed WIMP mass
 \cite{DMDDbg-f1v}:
 once the reconstructed WIMP mass is
 over-/underestimated from the real value,
 a transformation constant from $Q$ to $v$
 will thus be under-/overestimated.
 Consequently,
 the reconstructed $v$ points
 will be smaller/larger than the true values.

 In contrast to the first effect
 discussed in the previous subsection,
 this second effect
 draws the reconstructed WIMP velocity distribution
 (more strongly) to the opposite directions.
 Nevertheless,
 with an $\sim$ 5\% -- 10\% background ratio
 in the analyzed data sets of $\sim$ 500 total events,
 one could in principle still reconstruct
 the one--dimensional WIMP velocity distribution function
 with an $\sim +7\%$
 (for 25 GeV WIMPs, 10\% backgrounds)
 -- $\sim +16\%$
 (for 250 GeV WIMPs, 5\% backgrounds)
 deviation.
 If $\mchi~\lsim$ 100 GeV,
 the maximal acceptable background ratio
 could even be as large as $\sim$ 20\%
 with a deviation of only $\sim +9\%$.
\section{Summary}
 In this article
 we reexamine the data analysis methods
 introduced in Refs.~\cite{DMDDf1v, DMDDmchi}
 for determining the mass and one--dimensional
 velocity distribution function of halo Dark Matter particle
 from measured recoil energies of direct detection experiments directly,
 by taking into account small fractions of residue background events,
 which pass all discrimination criteria and
 then mix with other real WIMP--induced events
 in the analyzed data sets.

 Our simulations show that,
 with a background ratio of $\sim$ 5\% -- 10\%
 in data sets of $\sim$ 500 total events,
 the one--dimensional WIMP velocity distribution function
 can in principle be reconstructed
 with an $\sim -5\%$ -- $+20\%$ systematic deviation;
 whereas
 with a background ratio of even $\sim$ 10\% -- 20\%
 in data sets of only $\sim$ 50 total events,
 the 1$\sigma$ statistical uncertainty band
 of the reconstructed WIMP mass
 can cover the true value pretty well.
\subsubsection*{Acknowledgments}
%
 The author would like to thank
 the Physikalisches Institut der Universit\"at T\"ubingen
 for the technical support of the computational work
 demonstrated in this article.
 This work
 was partially supported by
 the National Science Council of R.O.C.~%
 under contract no.~NSC-98-2811-M-006-044
 as well as by
 the Focus Group on Cosmology and Particle Astrophysics,
 National Center of Theoretical Sciences, R.O.C..
\end{document}